\documentclass[peerreview,11pt]{IEEEtran}
\usepackage{graphicx}
\usepackage{fancyhdr}
\usepackage{makeidx}
\usepackage{amssymb,amsmath}
\usepackage{enumerate}
\newcommand{\bd}{\begin{document}}
\newcommand{\ed}{\end{document}}
\newcommand{\bc}{\begin{center}}
\newcommand{\ec}{\end{center}}
\newcommand{\vs}{\vspace}
\newcommand{\hs}{\hspace}
\newcommand{\beq}{\begin{equation}}
\newcommand{\eeq}{\end{equation}}
\newcommand{\beqs}{\begin{eqn*}}
\newcommand{\eeqs}{\end{eqn*}}
\newcommand{\bq}{\begin{quote}}
\newcommand{\eq}{\end{quote}}
\newcommand{\lb}{\linebreak}
\newcommand{\mb}{\makebox}
\newcommand{\fb}{\framebox}
\newcommand{\mc}{\multicolumn}
\newcommand{\ben}{\begin{enumerate}}
\newcommand{\een}{\end{enumerate}}
\newcommand{\bit}{\begin{itemize}}
\newcommand{\eit}{\end{itemize}}
\newcommand{\ov}{\overline}
\newcommand{\un}{\underline}
\newcommand{\lt}{\left}
\newcommand{\rt}{\right}
\newcommand{\ba}{\begin{array}}
\newcommand{\ea}{\end{array}}
\newcommand{\beqa}{\begin{eqnarray}}
\newcommand{\eeqa}{\end{eqnarray}}
\newcommand{\beqas}{\begin{eqnarray*}}
\newcommand{\eeqas}{\end{eqnarray*}}
\newcommand{\bfg}{\begin{figure}}
\newcommand{\efg}{\end{figure}}
\newcommand{\pad}{\partial}
\newcommand{\nn}{\nonumber}
\newcommand{\la}{\leftarrow}
\newcommand{\ra}{\rightarrow}
\newcommand{\lgla}{\longleftarrow}
\newcommand{\lgra}{\longrightarrow}
\newcommand{\La}{\Leftarrow}
\newcommand{\Ra}{\Rightarrow}
\newcommand{\Lra}{\Leftrightarrow}
\newcommand{\Lgla}{\Longleftarrow}
\newcommand{\Lgra}{\Longrightarrow}
\renewcommand{\a}{\alpha}
\renewcommand{\b}{\beta}
\newcommand{\g}{\gamma}
\newcommand{\G}{\Gamma}
\renewcommand{\d}{\delta}
\newcommand{\D}{\Delta}
\newcommand{\e}{\epsilon}
\newcommand{\eps}{\epsilon}
\newcommand{\s}{\sigma}
\renewcommand{\l}{\lamda}
\newcommand{\m}{\mu}
\newcommand{\n}{\nu}
\renewcommand{\S}{\Sigma}
\newcommand{\p}{\pi}
\newcommand{\om}{\omega}
\newcommand{\Om}{\Omega}
\newcommand{\tri}{\triangle}
\newcommand{\ti}{\times}
\newcommand{\f}{\frac}
\newcommand{\ds}{\displaystyle}
\newcommand{\bm}[1]{\mb{{\boldmath $#1$}}}
\newcommand{\alter}[2]{\lt\{ \ba{ll}#1 \\ #2 \ea \rt.}
\newcommand{\alt}[4]{\lt\{ \ba{ll}#1 & \mb{if \, \,}#2 \\ #3 & \mb{}#4 \ea
    \rt.}
\newcommand{\altn}[4]{\lt\{ \ba{rl}#1 & \mb{if \, \,}#2 \\ #3 & \mb{}#4 \ea
    \rt.}
\newcommand{\altif}[4]{\lt\{ \ba{ll}#1 & \mb{if \, \,}#2 \\ #3 &
\mb{if \, \,}#4 \ea \rt.}
\newcommand{\altnif}[4]{\lt\{ \ba{rl}#1 & \mb{if \, \,}#2 \\ #3 &
\mb{if \, \,}#4 \ea \rt.}
\newcounter{algc}
\newcounter{romc}
\newcounter{Alphc}
\newcommand{\bl}{\begin{list}{{\it Step} ~\arabic{algc}~:} {\usecounter{algc}
                \setlength{\topsep}{0pt} \setlength{\itemsep}{0pt}}}
\newcommand{\el}{\end{list}}
\newcommand{\blr}{\begin{list}{~\roman{romc}~:} {\usecounter{romc}
                \setlength{\topsep}{0pt} \setlength{\itemsep}{0pt}}}
\newcommand{\elr}{\end{list}}
\newcommand{\bla}{\begin{list}{~\Alph{Alphc}~:} {\usecounter{Alphc}
                \setlength{\topsep}{0pt} \setlength{\itemsep}{0pt}}}
\newcommand{\ela}{\end{list}}

\newtheorem{theorem}{Theorem}
\begin{document}
\title{Effects of Parasitics and Interface Traps On Ballistic Nanowire FET In
The Ultimate Quantum Capacitance Limit}
\author{Kausik Majumdar$^{*}$, Navakanta Bhat, Prashant Majhi$^{\dag}$ and Raj Jammy$^{\dag\dag}$\\
Department of Electrical Communication Engineering and Center of Excellence in Nanoelectronics,\\
Indian Institute of Science, Bangalore-560012, India\\
$^{\dag}$Intel Assignee at Sematech International, 2706 Montopolis
Dr, Austin, TX-78741, US\\
$^{\dag\dag}$Sematech International, 2706 Montopolis Dr, Austin, TX-78741,
US\\
$^*$Corresponding author, Email: kausik@ece.iisc.ernet.in}
\maketitle
{\abstract  In this paper, we focus on the performance of a nanowire
Field Effect Transistor (FET) in the Ultimate Quantum Capacitance
Limit (UQCL) (where only one subband is occupied) in the presence of
interface traps ($D_{it}$), parasitic capacitance ($C_L$) and
source/drain series resistance ($R_{s,d}$) using a ballistic
transport model and compare the performance with its Classical
Capacitance Limit (CCL) counterpart. We discuss four different
aspects relevant to the present scenario, namely, (i) gate voltage
dependent capacitance, (ii) saturation of the drain current, (iii)
the subthreshold slope and (iv) the scaling performance. To gain
physical insights into these effects, we also develop a set of
semi-analytical equations. The key observations are: (1) A strongly
energy-quantized nanowire shows non-monotonic multiple peak C-V
characteristics due to discrete contributions from individual
subbands; (2) The ballistic drain current saturates better in the
UQCL compared to CCL, both in presence and absence of $D_{it}$ and
$R_{s,d}$; (3) The subthreshold slope does not suffer any relative
degradation in the UQCL compared to CCL, even with $D_{it}$ and
$R_{s,d}$; (4) UQCL scaling outperforms CCL in the ideal condition;
(5) UQCL scaling is more immune to $R_{s,d}$, but presence of
$D_{it}$ and $C_L$ significantly degrades scaling advantages in the
UQCL.}\\

\newpage
\section{Introduction}
As we scale down the lateral as well as the longitudinal dimensions
of the channel of a non-planar transistor and replace the Silicon
channel by a so-called `high-mobility' (or low effective mass)
material, we start reaching two limits. The first limit is an
electrostatic one, termed as {\it `Quantum Capacitance Limit' (QCL)}
\cite{sl87}-\cite{kra08}. The strong energy quantization due to the
geometrical confinement in a multi-gated structure and the low
density of states due to small effective mass of the channel
material causes a small `quantum capacitance' $C_q$ to be in series
with the gate oxide capacitance $C_{ox}$. The small quantum
capacitance dominating the total gate capacitance results in a
number of interesting effects in the transistor characteristics
\cite{fwys07}-\cite{kra08}. There have been a number of experimental
efforts as well, in a variety of systems to capture the effect of
such quantum capacitance \cite{idk06,rna08}. The other limit, so
called {\it `Ballistic Transport Limit' (BTL)} is transport related
which results from the scaling of the channel length and the
relatively large mean free path of high mobility channel materials
\cite{kn94}-\cite{rkl05}. Scaling of multi-gate transistors thus
leads to a regime where the transistor is expected to be operating
in both the limits.

In this work, we analyze a Gate-All-Around nanowire transistor
operating in such limiting conditions. There have been a few reports
in the recent past where the performance of such a transistor has
been evaluated in the QCL in an ideal condition and compared with
the classical capacitance limit (CCL) where the gate oxide is
dominant \cite{kra08,kl08,kl09}. However, the effects of device
non-idealities on transistor performance become extremely important
in this regime. The presence of parasitic capacitance and interface
traps significantly reduces the fraction of the useful (mobile)
charge in the total switching charge in the strong quantum
capacitance limit. On the other hand, strong energy quantization
increases the relative channel resistance in this limit which in
turn improves immunity of the transistor towards the source/drain
series resistance. Thus, the combined effects of these
non-idealities are expected to play a significant role in the
transistor characteristics in the strong quantum capacitance limit
and are addressed in detail in the present work.

The paper is organized as follows: we describe the simulation model
used in this work in sec. \ref{sec:model}. In sec. \ref{sec:qcl}, we
start with a formal definition of the Ultimate Quantum Capacitance
Limit (UQCL), followed by a detailed analysis of the C-V
characteristics of a nanowire transistor in this regime in presence
of the interface traps. Sec. \ref{sec:analysis} presents a
comparative analysis between UQCL and CCL regime of operation on (i)
the saturation of drain current and (ii) the subthreshold slope,
both in the presence and absence of different device non-idealities.
In the same section, we also analyze the scaling performance of the
transistor in such limits. The performance benchmarking procedure
that we follow in this work is based on the criteria proposed in
\cite{rc05}. This is followed by the discussion on the effects of
the above said non-idealities on the scaling performance in sec.
\ref{sec:parasitics}. We demonstrate that the device non-idealities
should be carefully taken into account for realistic performance
evaluation in a nanowire transistor in the UQCL regime. Finally, we
conclude the paper in sec. \ref{sec:conclude}.
\section{A Ballistic Nanowire FET Model}\label{sec:model}
Here we use a FET model, which is a variation of the {\it
`Top-of-the-Barrier'} model described in {\cite{rgdl03}}, for a
gate-all-around (GAA) square nanowire of width $W$, schematically
shown in Fig. \ref{fig:nanowire}. We assume a parabolic
bandstructure of the nanowire described by an isotropic effective
mass $m^*$.

First, a hypothetical nanowire FET is assumed where the top of the
source to channel barrier at $x$=$x_0$ is physically far off from
the source and the drain eliminating any potential coupling from the
source or the drain. Then, the gate voltage ($V_g$) governed 2-D
potential profile $\phi_0(y,z)$ and the carrier density profile
$N_0(y,z)$ are obtained at $x$=$x_0$ plane of the hypothetical FET
using coupled 2-D Schrodinger-Poisson equations: \beq\label{eq:sch}
\left(\frac{\hbar^2}{2m^*}\frac{\partial^2}{\partial y^2} +
\frac{\hbar^2}{2m^*}\frac{\partial^2}{\partial z^2} + q\phi_0(y,z) +
E_i^0\right)\psi_i^0(y,z) = 0 \eeq and \beq\label{eq:pois}
\frac{\partial^2 \phi_0(y,z)}{\partial y^2} + \frac{\partial^2
\phi_0(y,z)}{\partial z^2} = \frac{q}{\epsilon}N_0(y,z) \eeq Here
$E_i^0$ is the energy minimum of the $i^{th}$ subband, $q$ and
$\epsilon$ are the electronic charge and dielectric constant of the
channel respectively. Assuming a ballistic channel, the carriers
with $+$$k$ and $-$$k$ states are in equilibrium with the chemical
potentials of the source ($\mu_s$) and the drain ($\mu_d$)
respectively with $\mu_d = \mu_s - qV_d$. Thus, \beq\label{eq:n}
N_0(y,z) = N_s(y,z) + N_d(y,z) \eeq where \beq\label{eq:nsd}
N_{s,d}(y,z) = \frac{1}{2}\sum_i\left[\int_{E_i^0}^\infty
D_i(E)f_{s,d}(E)\vert \psi_i^0(y,z)\vert^2dE\right] \eeq $D_i(E)$ is
the 1D density of states (DOS) of the $i^{th}$ subband in the
channel given by \beq\label{eq:dos} D_i(E) =
\frac{1}{\pi\hbar}\sqrt{\frac{m^*}{2(E-E_i^0)}}S(E-E_i^0) \eeq where
$S$ is a step function. $f_{s,d}(E)$ are the Fermi-Dirac
probabilities which, at temperature $T$, are expressed as
\beq\label{eq:fsd} f_{s,d}(E) = \frac{1}{1 + e^{(E -
\mu_{s,d})/k_BT}} \eeq where $k_B$ is the Boltzmann constant. The
wavefunction $\psi_i^0(y,z)$ is assumed to be zero at the
channel-gate dielectric interface. Once self-consistency is achieved
among Eq. (\ref{eq:sch}), (\ref{eq:pois}) and (\ref{eq:n}),
$\phi_0(y,z)$ and $N_0(y,z)$ correspond to the solutions at the top
of the barrier for the hypothetical long channel ballistic FET.

Now, to obtain the characteristics of a realistic short channel ballistic nanowire FET,
we define two terminal voltage dependent coupling parameters: the source and drain coupling parameters
$\alpha_s(V_g,V_d)$ and $\alpha_d(V_g,V_d)$ respectively such that the actual potential distribution
at the top of the source barrier is given by
\begin{equation}\label{eq:Phi}
\phi(y,z) = \alpha_sV_s + \alpha_dV_d + (1 - \alpha_s -
\alpha_d)\phi_0(y,z)
\end{equation}
For a given set of device parameters, the actual values of
$\alpha_s$ and $\alpha_d$ can be extracted by fitting 3-D simulation
data \cite{rgdl03}. It is important to note that the final potential
$\phi(y,z)$ at the top of the barrier has been found using a
non-self-consistent perturbation to $\phi_0(y,z)$, which, as we will
see later, helps us to develop a set of semi-analytical equations to
gain insights into transistor characteristics. However, this
introduces small inaccuracies in the short channel terminal
characteristics where the drain can have significant contribution to
$\phi(y,z)$. However, improved gate coupling due to the small $m^*$
and strong geometrical confinement in the present context largely
improves the gate to channel coupling, forcing $\alpha_s$ and
$\alpha_d$ to be small validating our assumption in Eq.
\ref{eq:Phi}. Although the model is valid for any arbitrary
dependence of the coupling parameters on terminal voltages, to
simplify the problem, we further assume that $\alpha_s$ and
$\alpha_d$ are independent of terminal voltages. In the rest of the
paper, we assume a ``well-behaved" transistor \cite{rgdl03,fwy04},
and do not explicitly mention about the length of the transistor,
rather assume that the short channel effects are being captured by
the parameters $\alpha_s$ and $\alpha_d$.

Using this $\phi(y,z)$, new energy eigen values $E_i$ and carrier
density $N(y,z)$ are recalculated. Finally, the ballistic current is
obtained by Landaur's formula: \beq\label{eq:Id} I_d =
\frac{q}{\pi\hbar}\sum_i\left[\int_{E_i}^\infty (f_s(E) -
f_d(E))dE\right] \eeq The effect of series source and drain
resistance ($R_{s,d}$) on the drain current are included as a
posteriori effect \cite{ml08}. In this work, the transistor, even in
presence of interface traps, has been considered to be ballistic to
keep the focus on the relative performance between the UQCL and the
CCL, assuming that the increase in scattering from the traps has a
similar impact on carrier transport in both the regimes. The
numerical simulation method described in this section has been used
to generate all the results to follow in the rest of the paper.
\section{The Ultimate Quantum Capacitance Limit (UQCL) in a Nanowire FET}\label{sec:qcl}
\subsection{Definition}
Generally, a qualitative definition, namely $C_q << C_{ox}$ is used
to identify whether a transistor is in the QCL or not. To give a
more quantitative definition taking care of the infinitely long high
energy tail of the Fermi-Dirac probability, in the present work, we
choose to call a FET to be operating in the UQCL when more than
$99\%$ of the carriers populate the first subband. Note that, in the
present context of low $m^*$ and strong geometrical confinement in a
nanowire, this formal definition automatically meets $C_q <<
C_{ox}$. In Fig. \ref{fig:qcl_Vg} we plot the maximum $V_g$ allowed
as a function of $W$ and $m^*$ to keep at least $99\%$ of the
carriers in the first subband. Thus, for any operating point in the
($V_g$, $m^*$, $W$) space that lies below the indicated surface is
said to be operating in the UQCL. In the CCL regime, the operating
point is expected to be much above the surface where large number of
subbands contribute. We also define a region called {\it
`Quasi-QCL'} which essentially represents those points which are
just above the indicated surface in Fig. \ref{fig:qcl_Vg}. Limited
number of subbands contribute to the total carrier density in this
regime. In the rest of the paper, we perform a comparative study of
two sets of points in the ($V_g$, $m^*$, $W$) space which are chosen
carefully such that one of them ($V_g=0.6V, m^*=0.07m_0, W \leq
10$nm) is into or very close to the UQCL regime of operation whereas
the other ($V_g=0.6V, m^*=0.5m_0, W \geq 10$nm) is closer to the CCL
regime. Note that, in both the cases, we assume the same bandgap
($E_g$) of $0.74$eV to compare them under the same terminal bias
conditions. This does not allow the difference in bandgap hide the
insights that we are looking for. Though this is a theoretical
construct, we do see such examples in reality for semiconductors
that are relevant to MOSFET channel. For example, both
In$_{x}$Ga$_{1-x}$As with $x=0.53$ and Si$_{1-y}$Ge$_{y}$ with
$y\approx 0.85$ have similar bandgap of $\sim 0.74$eV, however they
show a wide difference in the electron effective masses
\cite{mmbj}-\cite{web}. From Fig. \ref{fig:qcl_Vg}, it is
understandable that for a given $V_g$, reducing the nanowire cross
section or the effective mass of the channel material will push the
operating condition deeper into UQCL.
\subsection{Gate Capacitance}
For an undoped nanowire, the total gate capacitance for the
structure in Fig. \ref{fig:nanowire} can be expressed as a summation
of the contributions from all the subbands as \beq C_g(V_g) =
\sum_iC_{gi}(V_g) \eeq where \beq\label{eq:Cg} C_{gi}(V_g) =
\frac{\partial Q_{gi}(V_g)}{\partial V_g} = q\left[\frac{\partial
}{\partial V_g}\int_{E_i^0}^\infty D_i(E)f(E)dE\right] \eeq $Q_{gi}$
is the total mobile charge contribution from the $i^{th}$ subband.
Using Eq. (\ref{eq:dos}) and writing in terms of Fermi integral, we
obtain \beq\label{eq:Cgf} C_g(V_g) =
\frac{q}{\pi\hbar}(2m^*k_BT)^{-1/2}\sum_i\frac{\partial }{\partial
V_g}\left[F_{-1/2}\left(\frac{\mu_s - E_i^0}{k_BT}\right)\right]
\eeq The numerically simulated C-V characteristics are shown in Fig.
\ref{fig:Cg}(a) and (b) for multi-gate nanowire in the two different
regimes. It is clearly observed that, close to the CCL (Fig.
\ref{fig:Cg}(b)), $C_g$ scales almost linearly with the number of
gates $N_G$. However, in the UQCL (Fig. \ref{fig:Cg}(a)), the
scaling is much slower with $N_G$ due to the small series quantum
capacitance $C_q$. Consequently, migrating from double-gate to
gate-all-around structure is expected to be less effective in the
UQCL as compared to CCL regime.

We clearly observe the non-monotonicity of the nanowire C-V
characteristics close to UQCL in Fig. \ref{fig:Cg}(a). This is
explained in Fig. \ref{fig:Cg}(c) with a {\it
`parallel-subband-capacitance'} concept using the fact that the
total mobile charge is a sum of the contributions from the
individual subbands. Elementary electrostatics leads us to an
equivalent capacitance model using $C_{ox}$ and individual subband
quantum capacitances $C_{qi}$, as shown in the inset of Fig.
\ref{fig:Cg}(c). Unlike 2-D and 3-D structures, in a 1-D nanowire,
the DOS of individual subbands falls with energy as $E^{-1/2}$ (Eq.
\ref{eq:dos}). Into the UQCL, where only one subband contributes,
the impact of the DOS is manifested as a drop in $C_g$ after certain
$V_g$. This is arising from the non-monotonic behavior of the
function $\frac{\partial}{\partial x}F_{-1/2}(x)$ in the Eq.
(\ref{eq:Cgf}). However, as we move away from UQCL, more number of
subbands start contributing, and since each of them has its own
threshold (Fig. \ref{fig:Cg}(c)), the total gate capacitance shows
multiple peaks. Thus, the number of peaks in these highly quantized
nanowires is a signature of the number of subbands contributing to
the total carrier density. However, as we go closer to CCL, as shown
in Fig. \ref{fig:Cg}(d), large number of closely spaced subbands
contributing to the total carrier density destroy the humps.

We would like to mention a subtle point here: the physical origin of
$C_q$ is different from $C_{ox}$. When $C_{ox}$ is small, the
coupling between the gate and the channel degrades. However, a
strongly confined, low $m^*$ channel, resulting in low $C_q$,
provides an excellent gate control allowing the channel to attain
almost uniformly the terminal gate voltage with negligible drop
across the oxide.
\subsection{Effect of Interface Traps}
The reduced gate capacitance in the UQCL increases the impact due to
the interface traps. The issue is aggravated by the fact that, to
date, no high mobility channel MOSFET has been reported with
excellent gate insulator interface \cite{rna08,ng08}. Thus, it
becomes essential to include the effects of interface traps in the
UQCL regime of operation. In this paper, we assume that the trap
density $D_{it}$ is distributed uniformly over the bandgap of the
channel material and hence the boundary condition of the normal
components of the displacement vectors at the channel-insulator
interface is changed to $D_{\perp}^{ch} - D_{\perp}^{ins}=\rho_{it}$
where $\rho_{it}=-qD_{it}\int_{-\Delta E}^{0}f(E)dE$. $\Delta E$ is
the relative shift in the position of the intrinsic level. In Fig.
\ref{fig:Cg_Dit}(a) and (b), we plot the capacitance-voltage
characteristics in presence of different $D_{it}$ for UQCL and CCL
regime of operation. As expected, for a given $D_{it}$, particularly
at high $D_{it}$, the characteristics are significantly altered, and
the effect is more at UQCL compared to CCL. To investigate the
effect on mobile charges, we also plot the mobile charge capacitance
$C_m$ in Fig. \ref{fig:Cg_Dit}(c) and (d) which we define as the
rate of change of channel charge due to carriers with respect to
gate bias. It clearly shows that with an increase in the $D_{it}$,
the curves are shifted towards the right along the $V_g$ axis
indicating an increase in the threshold voltage.
\section{Nanowire FET Characteristics In The UQCL}\label{sec:analysis}
\subsection{Drain Current Saturation}
Saturation of the drain current with increase in drain bias is an
important requirement for the transistor to be useful for VLSI
applications. For example, lack of saturation in the drain current
leads to poor transfer characteristics of a CMOS inverter reducing
the noise margin of an SRAM cell. In the following, we show that a
ballistic FET shows better saturation characteristics in the UQCL
regime as compared to the CCL.

Assuming $\beta = e^{qV_d/k_BT}$ and $G_c = \frac{q}{\pi\hbar}$, Eq.
(\ref{eq:Id}) reduces to \beq\label{eq:sat1} I_d =
G_c\sum_i\left[\int_{E_i}^\infty \left(\frac{1}{1 + \Gamma(E)} -
\frac{1}{1 + \beta \Gamma(E)}\right)dE\right] \eeq where $\Gamma(E)
= e^{(E - \mu_s)/k_BT}$. Clearly, in the saturation region,
$\beta\Gamma(E) >> 1$. Thus, Eq. (\ref{eq:sat1}) gives
\beq\label{eq:sat2} I_d =
G_c\left(\frac{\beta-1}{\beta}\right)\sum_i\left(\int_{E_i}^\infty
\frac{dE}{1+\Gamma(E)}\right) \eeq Integrating, \beq\label{eq:sat3}
I_d = G_ck_BT\left(\frac{\beta-1}{\beta}\right)\sum_iln\left(1 +
e^{\frac{\mu_s - E_i}{k_BT}}\right) \eeq Now the perturbation to the
original Hamiltonian in Eq. (\ref{eq:sch}) due to introduction of
the source and drain coupling in Eq. (\ref{eq:Phi}) is given by
\beq\label{eq:delH} \Delta H = -q(\phi - \phi_0) = q(-\alpha_sV_s -
\alpha_dV_d + (\alpha_s + \alpha_d)\phi_0) \eeq Using first order
perturbation theory, the correction to the energy is obtained as
\begin{equation}\label{eq:delE}
\Delta E_i = \langle\psi_i^0|\Delta H|\psi_i^0\rangle = -q\alpha_sV_s - q\alpha_dV_d +
q(\alpha_s + \alpha_d)\langle\psi_i^0|\phi_0|\psi_i^0\rangle
\end{equation}
Hence, assuming source is grounded,
\beq\label{eq:Ei} E_i = E_i^0 + \Delta E_i = E_i^0 - q\alpha_dV_d +
q(\alpha_s + \alpha_d)\bar{\phi_i^0} \eeq where $\bar{\phi_i^0} =
\langle\psi_i^0|\phi_0|\psi_i^0\rangle$. We should note that, both
$E_i^0$ and $\bar{\phi_i^0}$ are almost independent of $V_d$ for
relatively large $V_d$ in saturation region since the contribution
of $N_d(y,z)$ to top of the source barrier carrier density $N(y,z)$
becomes relatively negligible in Eq. (\ref{eq:n}). Thus, Eq.
(\ref{eq:sat3}) reduces to \beq\label{eq:Idf} I_d =
G_ck_BT\left(\frac{\beta-1}{\beta}\right)\sum_iln\left(1 +
\beta^{\alpha_d}e^{\theta_i}\right) \eeq where $\theta_i =
\frac{\mu_s - E_i^0 - q(\alpha_s + \alpha_d)\bar{\phi_0}}{k_BT}$.
Here, $\beta$ is the only parameter that is dependent on $V_d$.
Thus, we can find the slope of the drain current in saturation
region as
\begin{equation}\label{eq:slopeId}
\frac{\partial I_d}{\partial V_d} = G_cq\beta^{-1}\sum_iln(1+\beta^{\alpha_d}e^{\theta_i}) +
G_cq\alpha_d(1 - \beta^{-1})\sum_i\frac{1}{1+\beta^{-\alpha_d}e^{-\theta_i}}
\end{equation}
Few conclusions can be drawn from Eq. (\ref{eq:slopeId}): (1) If
drain coupling parameter $\alpha_d \approx 0$ (the case of a
ballistic long channel hypothetical FET), the second term is very
small and the first term goes to zero exponentially with increase in
$V_d$, leading to excellent saturation. This argument supports the
results obtained from numerical simulations with $\alpha_d=0$, as
shown in Fig. \ref{fig:Idsat}(a). (2) For nonzero $\alpha_d$, $I_d$
does not completely saturate. However, with stronger quantization,
the magnitude of $E_i^0$ is larger which in turn reduces $\theta_i$,
hence reducing the magnitude of both the terms in Eq.
(\ref{eq:slopeId}). Consequently, if we operate close to the UQCL, a
short channel ballistic FET will show better saturation
characteristics as compared to a FET operating close to CCL. Similar
conclusion can be drawn from the results of numerical simulations
shown in Fig. \ref{fig:Idsat}(a) for $\alpha_d$=0.04 and $D_{it}$=0.

The presence of finite $D_{it}$ screens the gate voltage reducing
the magnitude of $\theta_i$ which in turn degrades the saturation
characteristics. In presence of extremely large $D_{it}$, as shown
in Fig. \ref{fig:Idsat}(a), UQCL loses its saturation performance
benefit over CCL. In Fig. \ref{fig:Idsat}(b), we show the output
characteristics of the same devices, but with
$R_s$=$R_d$=200$\Omega$-$\mu$m and
$D_{it}=10^{12}$eV$^{-1}$cm$^{-2}$ where it is observed that UQCL
operation retains its performance benefit over CCL. This is due to
the fact that stronger quantization in the UQCL compared to CCL
increases the relative channel resistance, resulting in more
immunity from series resistance effect.

From Eq. (\ref{eq:Idf}), we can also find the transconductance as
\begin{equation}\label{eq:G}
G = \frac{\partial I_d}{\partial V_g} =
-G_c\left(\frac{\beta-1}{\beta}\right)\sum_i\left[\frac{1}{1+\beta^{-\alpha_d}e^{-\theta_i}}\times\right.
\left.\frac{\partial}{\partial V_g}\left(E_i^0 + q(\alpha_s+\alpha_d)\bar{\phi_i^0}\right)\right]
\end{equation}
The numerically computed
transconductance (not shown) was found to have improved marginally
in the UQCL compared to its CCL counterpart.
\subsection{Subthreshold Slope}
To obtain the subthreshold slope of such a nanowire FET, we find
from Eq. (\ref{eq:Idf}):
\begin{equation}
ln(I_d) = ln(G_ck_BT) + ln\left(\frac{\beta-1}{\beta}\right) +
ln\left[\sum_iln\left(1 +
\beta^{\alpha_d}e^{\theta_i}\right)\right]
\end{equation}
which gives
\begin{equation}
\frac{\partial ln(I_d)}{\partial V_g} = \left(\frac{-1}{k_BT}\right)\left(\frac{1}{\sum_i\left(ln(1+e^{\eta_i})\right)}\right)\times
\sum_i\left[\frac{e^{\eta_i}}{1+e^{\eta_i}}\frac{\partial}{\partial
V_g}\left(E_i^0 + q(\alpha_s+\alpha_d)\bar{\phi_i^0}\right)\right]
\end{equation}
\setlength{\arraycolsep}{5pt} where $\eta_i = \theta_i +
\alpha_d\frac{qV_d}{k_BT}$. Now, at subthreshold, coupled with the
low $V_d$ operation, $\eta_i < 0$ and $|\eta_i| >> 1$ which gives
$ln(1+e^{\eta_i})\simeq e^{\eta_i}$. Also, at subthreshold, both
$\frac{\partial \bar{\phi_i^0}}{\partial V_g}$ and $\frac{\partial
E_i^0}{\partial V_g}$ are nearly constant, and let us call them
$\kappa_1$ and $-q\kappa_2$. Thus we obtain, \beq \frac{\partial
ln(I_d)}{\partial V_g} \simeq \frac{q}{k_BT} \left(\kappa_2 -
(\alpha_s+\alpha_d)\kappa_1\right) \eeq Hence, the subthreshold
slope is \beq\label{eq:SS} S = ln(10)\frac{\partial V_g}{\partial
ln(I_d)} =
\left(ln(10)\frac{k_BT}{q}\right)\times\frac{1}{\kappa_2-(\alpha_s+\alpha_d)\kappa_1}
\eeq In the case of infinite channel ($\alpha_s$=$\alpha_d$=0) with
no $D_{it}$, putting $\kappa_2$ as $1$, the limit of subthreshold
slope reduces to expected $ln(10)\frac{k_BT}{q}$ (=$60$mV/decade).
Note that, both $\kappa_1$ and $\kappa_2$ are relatively insensitive
to the region of operation (UQCL or CCL) resulting in similar
subthreshold slopes, as shown in Fig. \ref{fig:Idsat}(c). The same
trend remains even in the presence of moderately large $D_{it}$ and
hence both UQCL and CCL suffer from similar degradation in
subthreshold slope in presence of $D_{it}$. When $\alpha_s$ and
$\alpha_d$ are small, from Eq. \ref{eq:SS}, we can express $S$ in
mV/decade as \beq S \simeq \frac{60}{\kappa_2} \times \left(1 +
(\alpha_s+\alpha_d)\frac{\kappa_1}{\kappa_2}\right) \eeq which
explains the almost linear degradation of subthreshold slope with
$\alpha_d$ at a given $D_{it}$ in Fig. \ref{fig:Idsat}(c).
\subsection{Scaling and Performance}
\begin{figure}
\bc
\includegraphics[scale=0.7]{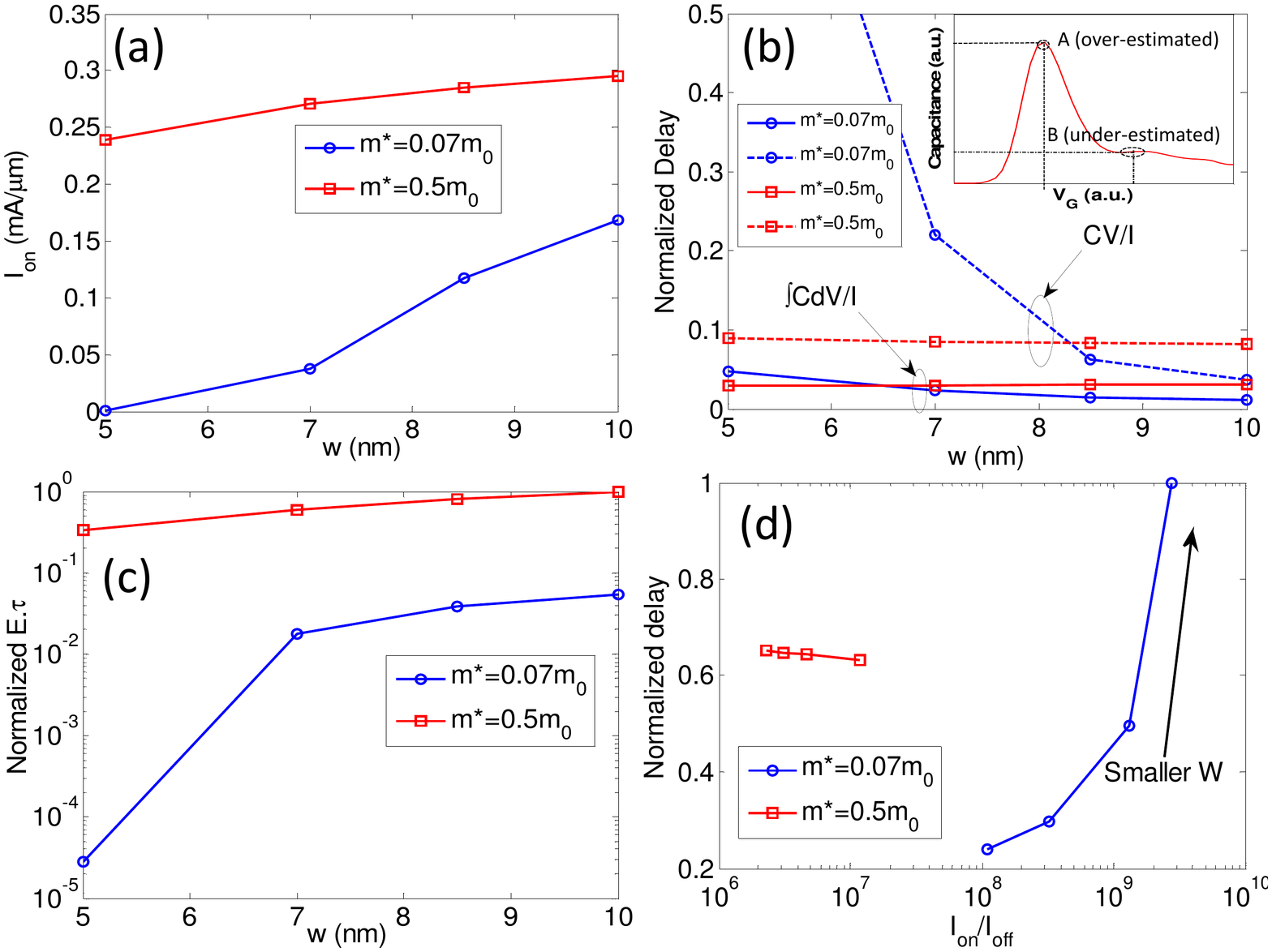}
\caption{}\label{fig:perfm} \ec
\end{figure}

In this section, we will compare the performance of the ballistic
nanowire FET between UQCL and CCL. Due to stronger quantization, the
ON current, normalized with the perimeter of the nanowire, degrades
in the vicinity of UQCL, which is shown in Fig. \ref{fig:perfm}(a).
It can also be observed that in this regime, the normalized ON
current is more sensitive to the nanowire dimension ($W$), which can
be a potential cause to add variability to the device.

Let us now present the relative performance of UQCL and CCL in the
light of the performance metrics proposed in \cite{rc05}. It has
been pointed out in \cite{kra08} that there is no scaling
disadvantage in CV/I metric while operating in the UQCL, for both
ballistic as well as non-ballistic transport. However, following the
discussion of C-V characteristics in sec. \ref{sec:qcl}, we wish to
point out that CV/I may not be able to reflect the intrinsic gate
delay accurately in the UQCL due to the non-monotonic dependence of
$C_g$ on $V_g$. This is explained in the inset of Fig.
\ref{fig:perfm}(b) where only one subband contributes to the total
mobile charge pushing the transistor deep into UQCL. As shown, CV/I
will over estimate the delay at point A whereas at point B, it will
under estimate due to wrong computation of the total charge as
indicated by the dotted rectangles. We thus take $\frac{\int
C_gdV_g}{I_d}$ as the delay metric in this work which represents the
gate delay more accurately by taking care of the details of the C-V
curve.

Using Eqs. (\ref{eq:Cgf}) and (\ref{eq:Idf}), we find the intrinsic
gate delay as
\begin{equation}\label{eq:delay}
\tau = \frac{\int C_gdV_g}{I_d} = (2m^*)^{-1/2}(k_BT)^{-3/2}\left(\frac{\beta}{\beta-1}\right)\times
\left[\frac{\sum_i
F_{-1/2}(\frac{\mu_s-E_i^0}{k_BT})}{\sum_iF_0(\frac{\mu_s-E_i}{k_BT})}\right]
\end{equation}

Fig. \ref{fig:perfm}(b) shows the numerically computed delay as a
function of the nanowire dimension from two different metrics. It is
found that in the UQCL, CV/I significantly over estimates the delay.
However, it becomes closer to the $\frac{\int C_gdV_g}{I_d}$ metric
as we move out of the UQCL regime (larger $W$). Note that, in a very
narrow nanowire, lower $m^*$ channel material shows higher gate
delay due to strong ON current degradation.

Energy-delay product, which has been obtained as $E.\tau = (\int
QdV).(\int CdV)/I_d$, is found to be very impressive at UQCL (Fig.
\ref{fig:perfm}(c)) as compared to its CCL counter part. This is
because the transistor at UQCL operates at significantly lower
switching charge. The normalized delay versus $I_{on}/I_{off}$ has
been plotted in Fig. \ref{fig:perfm}(d) for both UQCL and CCL. It
can be clearly seen that at UQCL, the transistor operation can be
pushed towards the ideal right-bottom corner of the delay -
$I_{on}/I_{off}$ space. However, in the extremely deep UQCL (smaller
$W$ in the plot), the delay performance is adversely impacted due to
excessive degradation of ON current.
\section{Scaling And Performance In Presence Of Non-idealities}\label{sec:parasitics}
The fact that a transistor in its UQCL regime of operation switches
relatively small amount of mobile charge, it is prone to performance
degradation in presence of parasitic capacitances and traps since
the fraction of the useful charge (mobile charge that drives the
drain current) reduces significantly in the total switching charge.
Here we separately discuss the effects of the interface traps,
parasitic capacitance and source/drain series resistance, and
finally comment on the combined effect of all of them.

\begin{figure}
\bc
\includegraphics[scale=0.7]{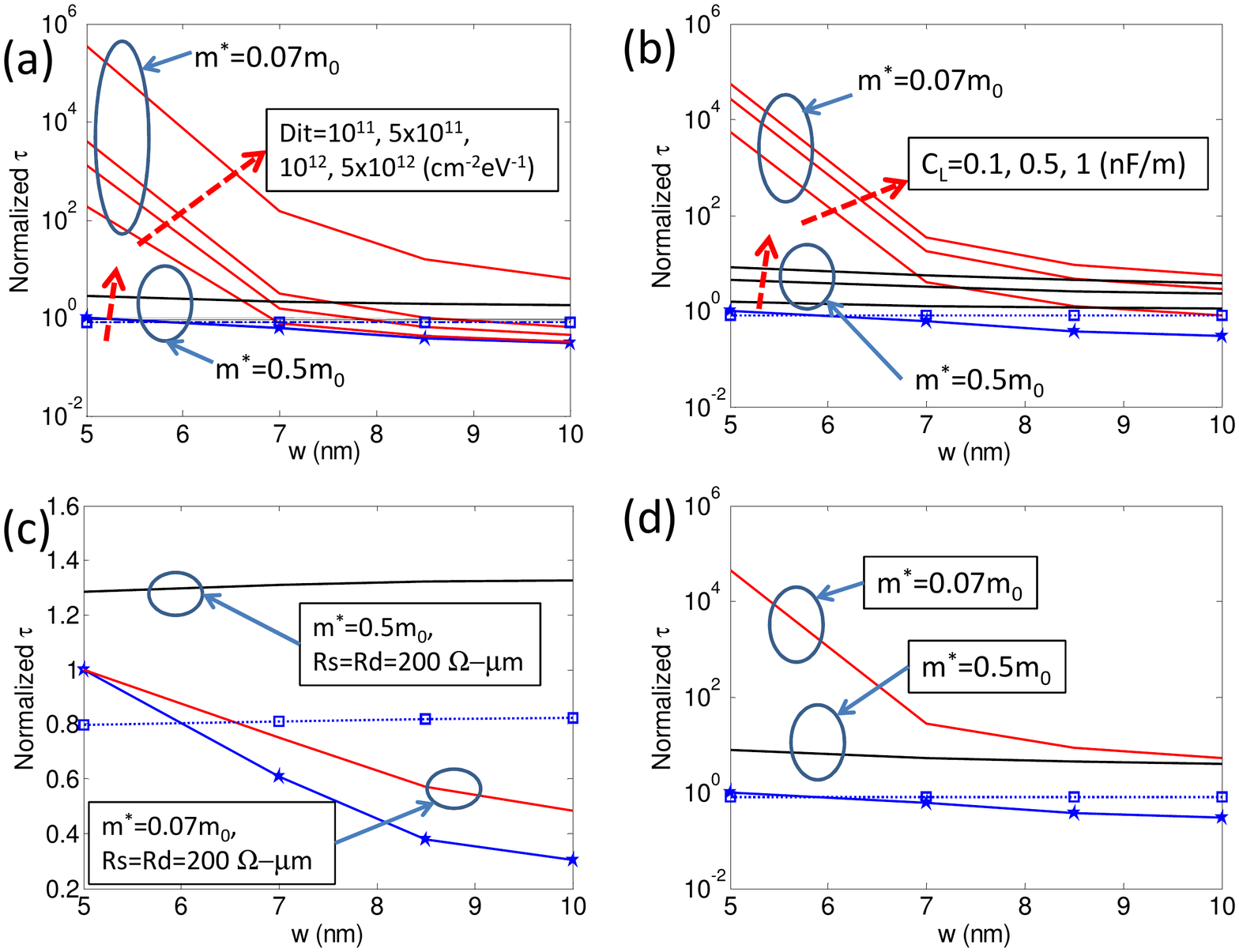}
\caption{}\label{fig:Rsd} \ec
\end{figure}

Fig. \ref{fig:Rsd}(a) shows the effect of $D_{it}$ on intrinsic
delay of the FET, for two different $m^*$. Increase in $D_{it}$
increases the threshold voltage reducing the ON current and also
increases the unused charge at the dielectric interface, degrading
the gate delay. As expected, we observe that the delay in the larger
$m^*$ channel is less sensitive to $D_{it}$ compared to the lower
$m^*$ and there is a severe degradation of delay at the narrower
dimension for low $m^*$ material. However, at comparatively larger
dimension ($W$~10nm), it is still possible to retain the relative
speed advantage of the low $m^*$ case, even with significant
$D_{it}$. We observe similar trends when we have parasitic
capacitance as the only source of non-ideality, as shown in Fig.
\ref{fig:Rsd}(b). This is again due to the fact that when we are
deep into quantum capacitance limit, presence of parasitic
capacitance significantly reduces the ratio of mobile charge to
total charge that is switched. However, it is interesting to note
that the relative degradation of the intrinsic gate delay in the
quantum capacitance limit is less compared to the classical
capacitance limit in the presence of series source/drain resistance
($R_{s,d}$), which is shown in Fig. \ref{fig:Rsd}(c). This arises
due to the increased channel resistance in the UQCL due to stronger
energy quantization, improving the immunity from parasitic series
resistance effect. In fact, deep into the quantum capacitance limit
(lower $m^*$ and lower $W$), presence of series resistance does not
at all alter the intrinsic gate delay, as shown in Fig.
\ref{fig:Rsd}(c). Finally, \ref{fig:Rsd}(d) shows the combined
effect of all the three non-idealities together which clearly shows
a negative impact on the speed advantages that UQCL can have over
CCL in the ideal scenario.

\begin{figure}
\bc
\includegraphics[scale=0.7]{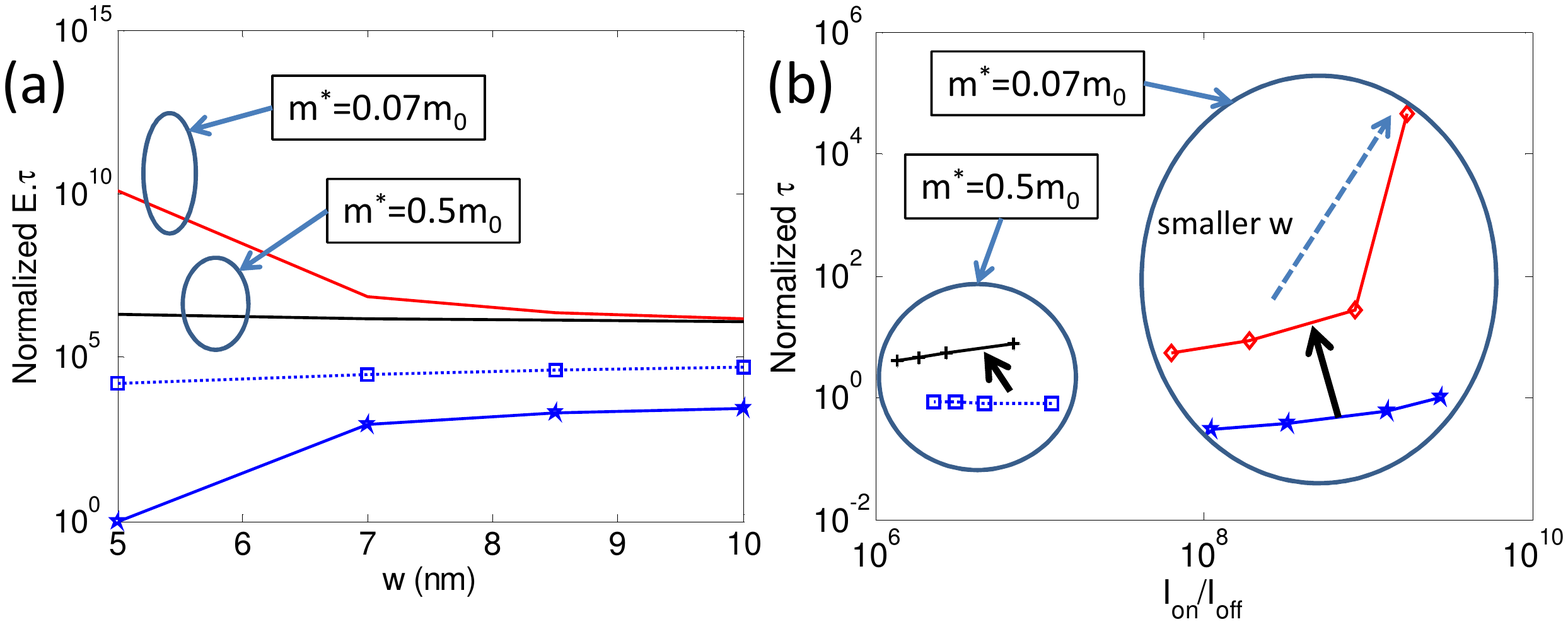}
\vs{-1.5in} \caption{}\label{fig:Etau} \ec
\end{figure}

In Fig. \ref{fig:Etau}(a), we present the energy-delay product in
presence of similar non-ideal conditions showing a significant
relative degradation in the UQCL. However, as shown in Fig.
\ref{fig:Etau}(b), even in the presence of the non-idealities, UQCL
operation manages to offer significantly higher ON to OFF current
ratio at comparable gate delays if we choose $W$ appropriately. We
conclude that the extremely deep UQCL may not be the optimum design
for a ballistic nanowire FET in presence of parasitics. At the same
time, to retain the OFF performance advantage, better drain current
saturation and immunity from series resistance, designing much away
from UQCL is also not desirable. The {\it `Quasi-QCL'} regime (where
few subbands contribute) with low $m^*$ and moderately large $W$,
can provide the optimum design to achieve both ON and OFF state
performance.
\section{Conclusion}\label{sec:conclude}
To conclude, a ballistic nanowire FET model has been proposed in the
work, which is a variation of the \emph{`Top-of-the-barrier'} model,
to analyze the transistor characteristics. The UQCL and CCL regimes
have been formally defined in the ($V_g$, $m^*$, $W$) design space
based on the subband occupation. The non-monotonic C-V
characteristics close to UQCL regime has been explained using a {\it
`parallel-subband-capacitance'} model. The small quantum capacitance
has been found to play a critical role in this regime in presence of
interface traps. The saturation characteristics of the drain current
are found to improve in the UQCL as compared to CCL regime, both in
presence and absence of parasitic resistance and interface traps. It
has also been found that the subthreshold slope in UQCL is similar
to CCL, even in the presence of interface traps. In the ideal
condition, the scaling performance at UQCL regime has been shown to
outperform its CCL counterpart. The UQCL operation has been found to
be more immune from series resistance effect compared to CCL,
whereas the presence of interface traps and parasitic capacitance
are shown to diminish the relative performance advantages of the
UQCL operation significantly. In presence of the combined effects of
all these parasitics, the UQCL operation retains its
$I_{on}/I_{off}$ advantage at comparable gate delay. A careful
design in the {\it `Quasi-QCL'} regime with low effective mass and
moderate nanowire width is required to obtain optimum ON and OFF
performance.
{\section*{Acknowledgement} K.M. and N.B. would like to
sincerely acknowledge the support from the Ministry of Communication
and Information Technology (MCIT), Govt. of India, and the
Department of Science and Technology (DST), Govt. of India.}

\newpage
{\it {\bf FIGURE CAPTIONS:}}

Fig. 1: a): Schematic diagram of a gate-all-around nanowire
FET. (b): Locus of the conduction band minimum from source to drain
with the top of the barrier height at $x$=$x_0$.

Fig. 2: UQCL regime of operation in the ($V_g, m^*, W$) space: More
than $99\%$ of the carriers populate only the first subband at any
operating point below the indicated surface and hence operates in
the UQCL, whereas multiple subbands are occupied at points above the
surface. The CCL regime, where a large number of subbands
contribute, is far above the surface.

Fig. 3: (a): Total gate capacitance $C_g$ for $m^*$=0.07$m_0$,
$W$=10nm and EOT=1nm ($C_{ox}$=0.35$\times$$N_G$ in nF/m) shows
non-monotonic C-V characteristics. The number gates $N_G$ is 2, 3
and 4. (b): Same as (a) with $m^*$=0.5$m_0$. (c): Contribution from
individual subbands to the total capacitance causes multi-peak C-V
close to UQCL. The inset shows a {\it
`parallel-subband-capacitance'} model. (d): Large number of subbands
contribute to the total capacitance close to CCL resulting a smooth
monotonic characteristics.

Fig. 4: Total gate capacitance $C_g$ in presence of interface traps
for $W$=10nm with (a): $m^*$=0.07$m_0$ and (b): $m^*$=0.5$m_0$. The
$D_{it}$ is assumed to be $10^{10}, 10^{11}, 10^{12},
5\times10^{12}$ and $10^{13}$ eV$^{-1}$cm$^{-2}$. (c)-(d): The
corresponding mobile charge capacitance (rate of change of mobile
charge in the channel with $V_g$) in the two scenarios.

Fig. 5: (a): Output characteristics of a gate-all-around 10nm X 10nm
ballistic nanowire FET with EOT=1nm and $V_g$=0.6V for $\alpha_d$=0
and 0.04. The red and blue curves correspond to $m^*$ = 0.07$m_0$
and 0.5$m_0$ respectively. The circles and squares correspond to
$D_{it}$=0 and $5\times10^{12}$ eV$^{-1}$cm$^{-2}$ respectively.
(b): The output characteristics in presence of parasitic source and
drain resistance of 200$\Omega$-$\mu$m and
$D_{it}=10^{12}$eV$^{-1}$cm$^{-2}$. The red and blue curves
correspond to $m^*$ = 0.07$m_0$ and 0.5$m_0$ respectively. (c):
Degradation of subthreshold slope with $D_{it}$ concentration for
$\alpha_d$=0, 0.05, 0.1 and 0.15. The solid and dotted curves (which
almost coincide) represent $m^*$ = 0.07$m_0$ and 0.5$m_0$
respectively. $\alpha_s$=0.01 for all the curves.

Fig. 6: (a): ON current as a function of nanowire width $W$. (b):
Normalized intrinsic gate delay ($\tau$) computed from two different
metrics as a function of $W$ for $m^*$ = 0.07$m_0$ and 0.5$m_0$. In
the inset, The delay predicted by $CV/I$ is given by the area of the
rectangles shown by the dotted lines, which depends on the operating
condition and can either overestimate (at point A) or underestimate
(at point B) the actual delay given by $\int CdV/I$. (c): Lower
switching charge results in better energy-delay product ($E.\tau$)
at UQCL. (d): Delay versus $I_{on}/I_{off}$ for different $W$. In
all the plots, EOT=1nm, $\alpha_d$=0.05, $\alpha_s$=0.01, $V_d$=0.6V
and $V_g$=0.6V have been assumed.

Fig. 7: (a): Normalized gate delay in presence of (a): only $D_{it}$
with $R_{s,d}$=0 and $C_L$=0, (b): only $C_L$ with $D_{it}$=0 and
$R_{s,d}$=0, (c): only $R_{s,d}$ with $D_{it}$=0 and $C_L$=0 and (d)
$R_{s,d}$=200$\Omega$-$\mu$m,
$D_{it}$=$5\times10^{11}$cm$^{-2}$eV$^{-1}$ and $C_L$=0.5nF/m. In
all the plots, the blue lines with star and squares represent the
reference delay with zero non-ideality for $m^*$=0.07$m_0$ and
$m^*$=0.5$m_0$ respectively. EOT=1nm, $\alpha_d$=0.05,
$\alpha_s$=0.01, $V_d$=0.6V and $V_g$=0.6V have been assumed.

Fig. 8: (a): Energy-delay product and (b) gate delay versus
I$_{on}$/I$_{off}$ plot with $R_{s,d}$=200$\Omega$-$\mu$m,
$D_{it}$=$5\times10^{11}$cm$^{-2}$eV$^{-1}$ and $C_L$=0.5nF/m. In
both the plots, the blue lines with star and squares represent the
reference with zero non-ideality for $m^*$=0.07$m_0$ and
$m^*$=0.5$m_0$ respectively. EOT=1nm, $\alpha_d$=0.05,
$\alpha_s$=0.01, $V_d$=0.6V and $V_g$=0.6V have been assumed.

\newpage
\bfg[htbp!] \bc
\includegraphics[scale=0.65]{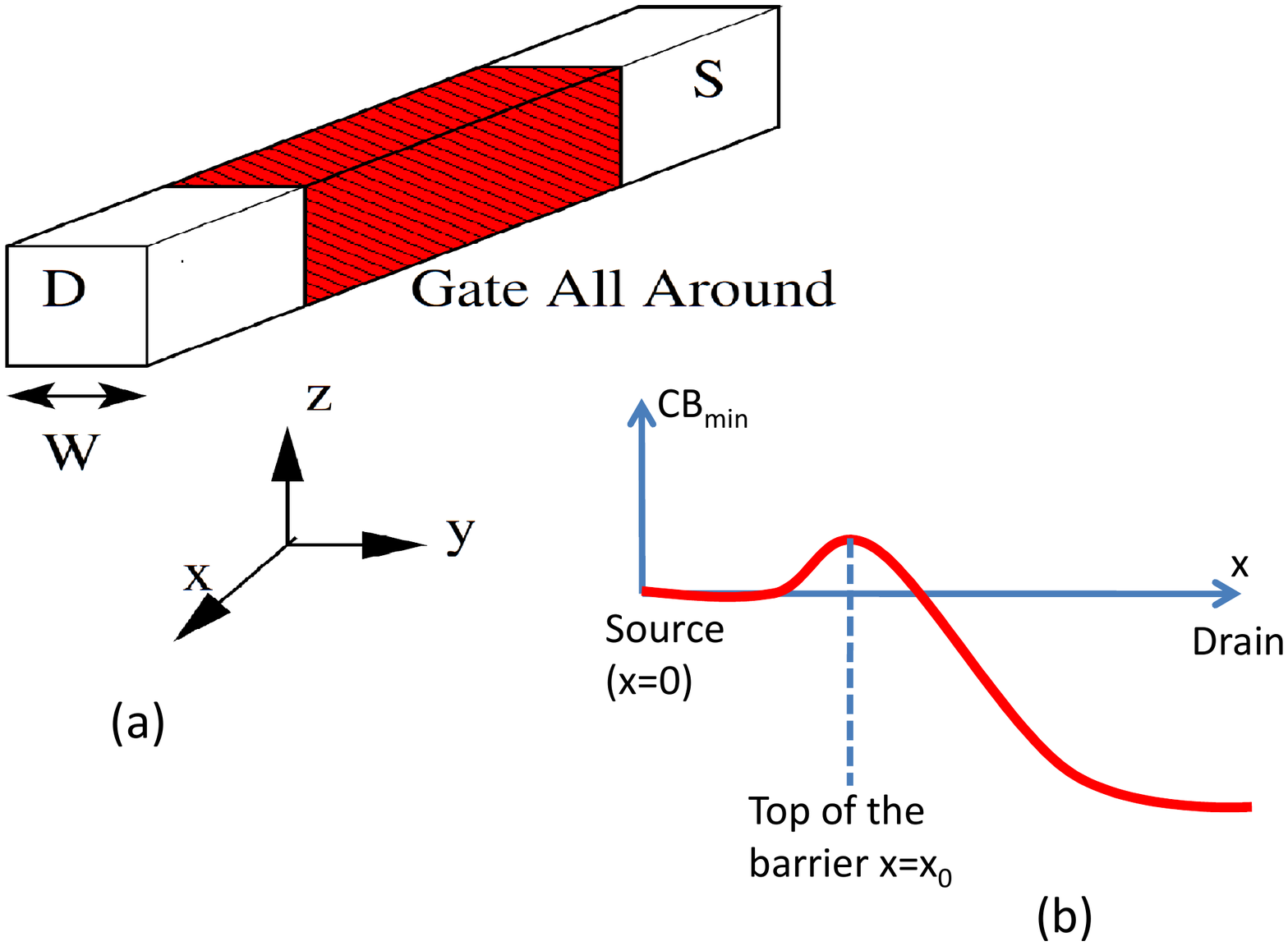}
\vs{-0.7in} \caption{}\label{fig:nanowire} \ec \efg

\newpage
\bfg[htbp!] \bc
\includegraphics[scale=0.7]{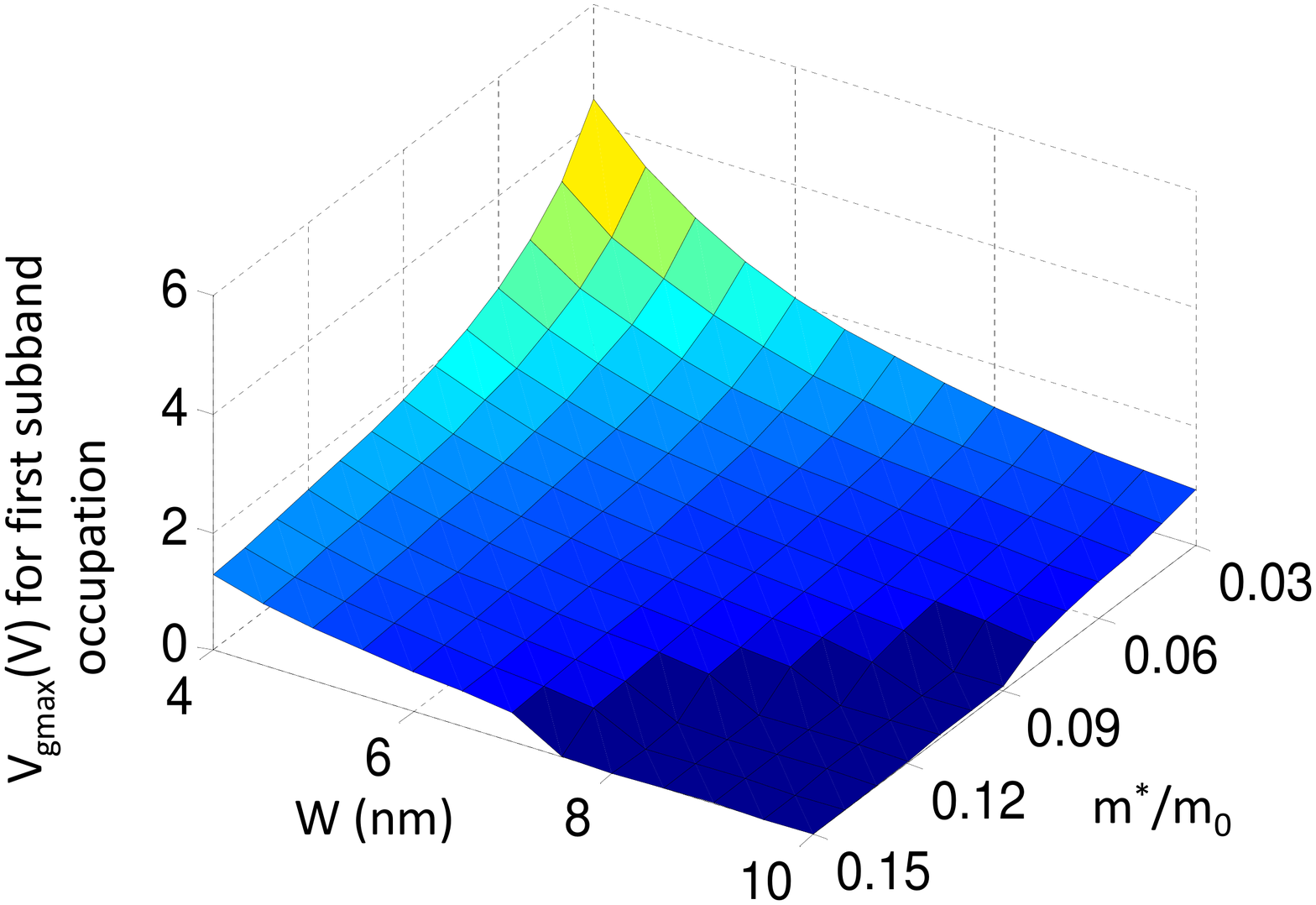}
\caption{}\label{fig:qcl_Vg} \ec \efg

\newpage
\bfg[htbp!] \bc
\includegraphics[scale=0.7]{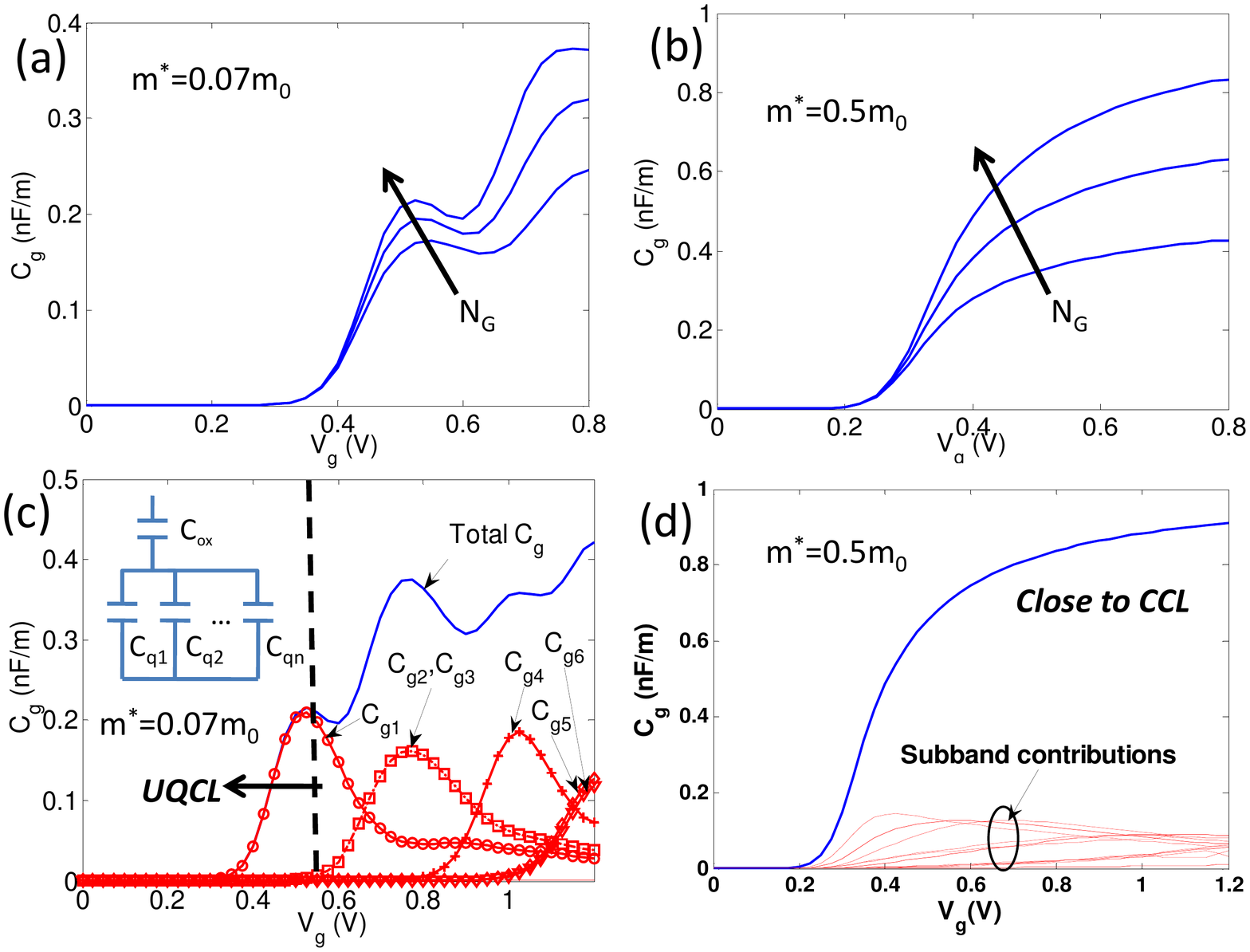}
\caption{}\label{fig:Cg} \ec \efg

\newpage
\bfg[htbp!] \bc
\includegraphics[scale=0.7]{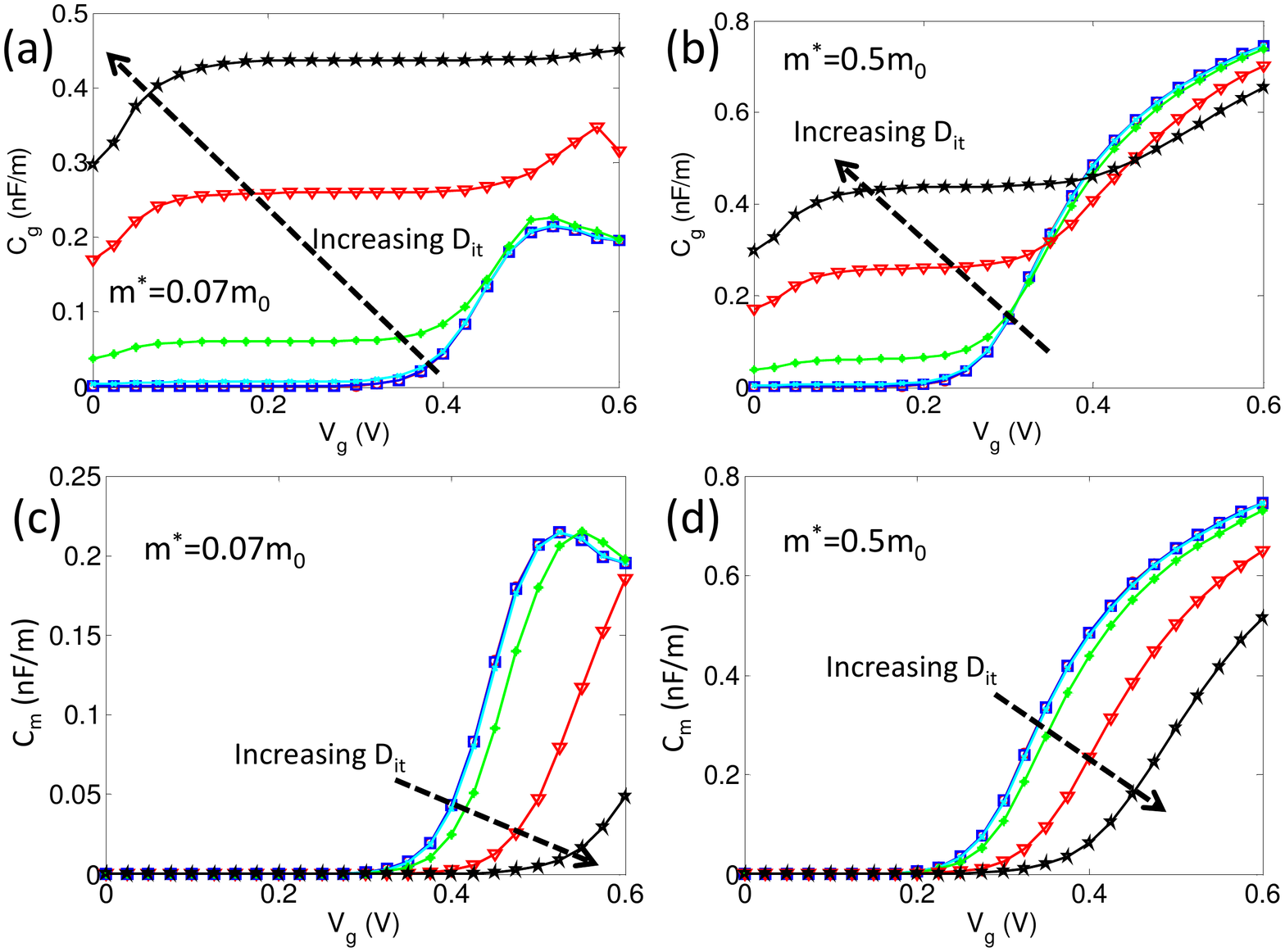}
\caption{}\label{fig:Cg_Dit} \ec \efg

\newpage
\bfg[htbp!] \bc \vs{-0.2in}
\includegraphics[scale=0.7]{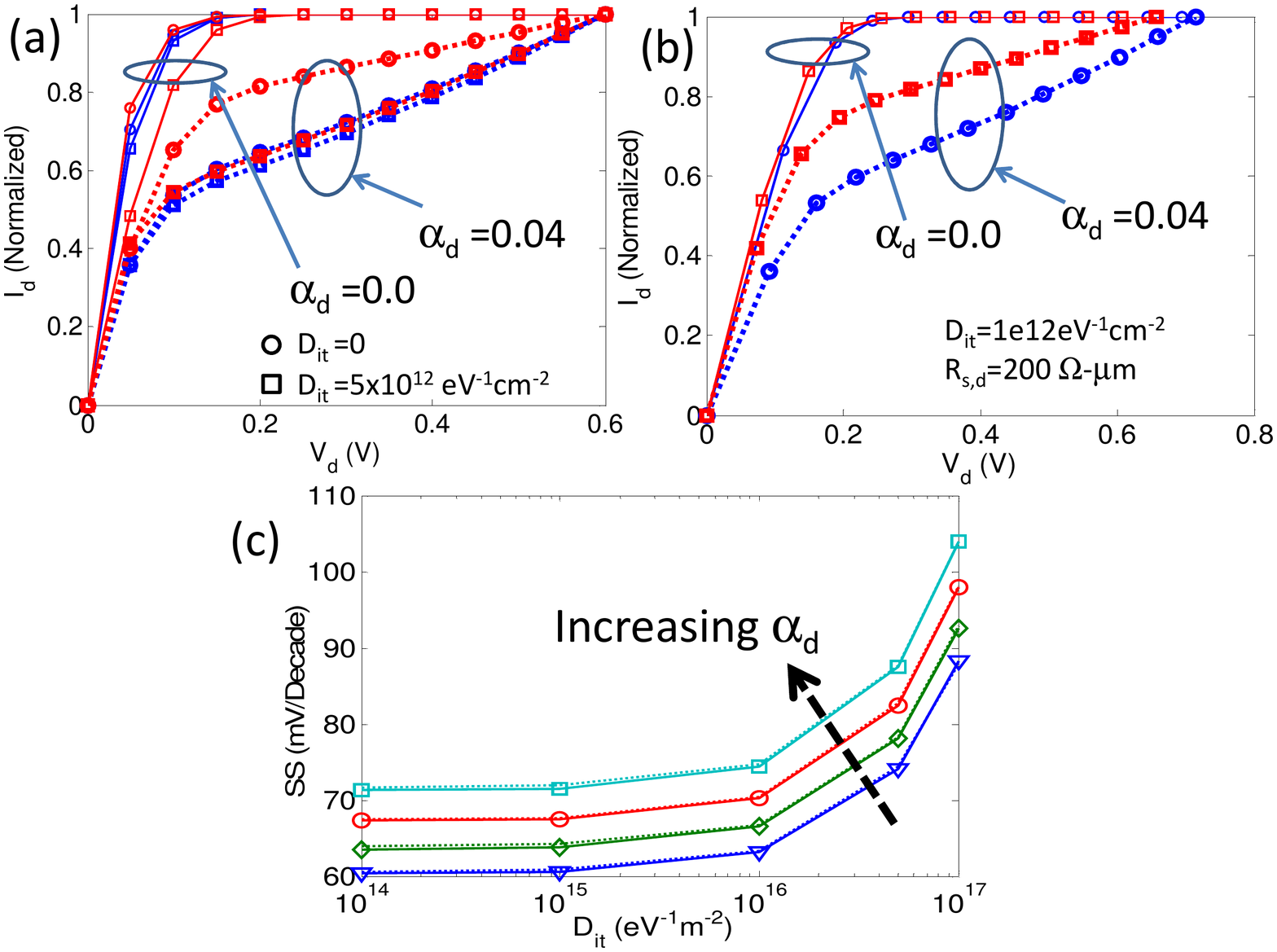}
\vs{-0.2in} \caption{}\label{fig:Idsat} \ec \efg

\newpage

\newpage

\newpage

\end{document}